\documentclass[acmtog]{acmart}

\usepackage{graphicx}
\usepackage{amsfonts}
\usepackage{amsmath}
\usepackage{booktabs}
\usepackage{xcolor}
\usepackage{multirow}
\usepackage[normalem]{ulem}
\usepackage[super]{nth}
\usepackage{cleveref}
\usepackage{colortbl}
\usepackage{xspace}

\acmJournal{TOG}

\acmSubmissionID{485}

\newcommand{\nbframe}{ H }
\newcommand{\nbframein}{ T }
\newcommand{\nbframeout}{ F }
\newcommand{\nbjoint}{ J }
\newcommand{\rotdim}{ Q }
\newcommand{\realnum}{ \mathbb{R} }
\newcommand{\rootpos}{ \textbf{O} }
\newcommand{\rootdelta}{ \textbf{V} }
\newcommand{\jointrot}{ \textbf{R} }
\newcommand{\motionmetric}{ \mathcal{M} }
\newcommand{\motionfeatin}{ \textbf{T} }
\newcommand{\motionfeatout}{ \textbf{F} }

\newcommand{\nbstage}{ S }
\newcommand{\idstage}{ s }
\newcommand{\downsamplef}{ r }
\newcommand{\upsample}{ \uparrow^{ \downsamplef} }

\newcommand{ \patchsize }{ p }
\newcommand{\skeleton}{ \textbf{B} }
\newcommand{\skeletalpart}{ \bar{\skeleton} }
\newcommand{\nbpart}{B}
\newcommand{\idpart}{b}
\newcommand{\partialmotion}{ \bar{\motionfeatin} }

\newcommand{\qset}{ X }
\newcommand{\kset}{ Y }
\newcommand{\dist}{ D }
\newcommand{\simi}{ \hat{\dist} }

\newcommand{\nbiter}{ E }

\newcommand{\contactlabel}{ \textbf{L} }

\newcommand{\exemplar}{ E }

\newcommand{\ncontact}{ C }

\setcitestyle{nosort,square}

\begin{document}

\title{Example-based Motion Synthesis via Generative Motion Matching
}

\author{Weiyu Li}
\email{weiyuli.cn@gmail.com}
\orcid{0000-0003-4500-4905}
\affiliation{%
  \institution{Shandong University}
  \country{China}
}
\authornote{Joint first authors}
\authornote{Work done during an internship at Tencent AI Lab}

\author{Xuelin Chen}
\email{xuelin.chen.3d@gmail.com}
\orcid{0009-0007-0158-9469}
\affiliation{%
  \institution{Tencent AI Lab}
  \country{China}
}
\authornotemark[1]
\authornote{Corresponding author}

\author{Peizhuo Li}
\email{peizhuo.li@inf.ethz.ch}
\orcid{0000-0001-9309-9967}
\affiliation{%
  \institution{ETH Zurich}
  \country{Switzerland}
}

\author{Olga Sorkine-Hornung}
\email{sorkine@inf.ethz.ch}
\orcid{0000-0002-8089-3974}
\affiliation{%
  \institution{ETH Zurich}
  \country{Switzerland}
}

\author{Baoquan Chen}
\email{baoquan@pku.edu.cn}
\orcid{0000-0003-4702-036X}
\affiliation{%
  \institution{Peking University}
  \country{China}
}

\renewcommand{\shortauthors}{Weiyu Li, Xuelin Chen, Peizhuo Li, Olga Sorkine-Hornung, and Baoquan Chen}
\newcommand{\name}{GenMM\xspace}

\begin{abstract}

We present \name, 
a generative model that ``mines'' as many diverse motions as possible from a single or few example sequences.
In stark contrast to existing data-driven methods,
which typically require long offline training time, are prone to visual artifacts, and tend to fail on large and complex skeletons,
\name inherits the training-free nature and the superior quality of the well-known \emph{Motion Matching} method.
\name can synthesize a high-quality motion within a fraction of a second, 
even with highly complex and large skeletal structures.
At the heart of our generative framework lies the generative motion matching module,
which utilizes the bidirectional visual similarity as a generative cost function to motion matching,
and operates in a multi-stage framework to progressively refine a random guess using exemplar motion matches.
In addition to diverse motion generation,
we show the versatility of our generative framework by extending it to a number of scenarios that are not possible with motion matching alone,
including motion completion, key frame-guided generation, infinite looping,
and motion reassembly.

\end{abstract}



\begin{teaserfigure}
\center
  \includegraphics[width=\textwidth]{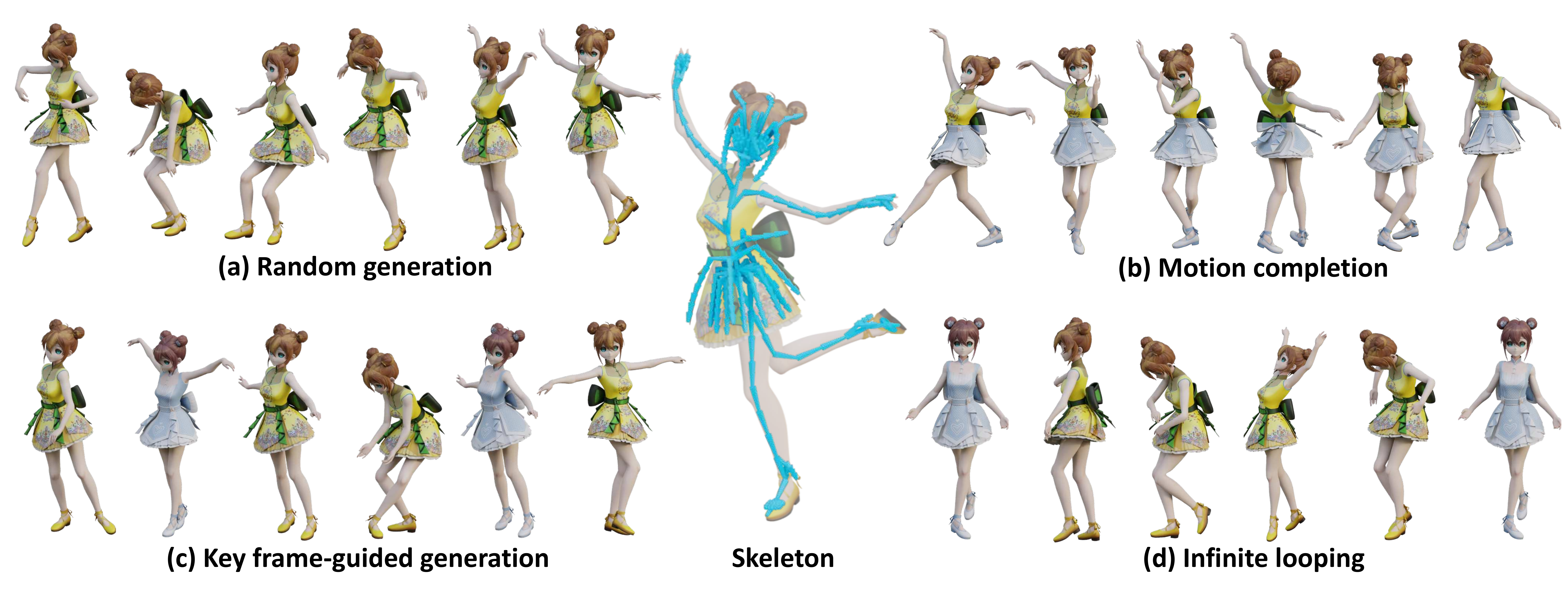}
  \caption{
  Our generative framework enables a variety of example-based motion synthesis tasks, that usually require long offline training for existing data-driven methods.
  Given a single or few examples,
  even with a highly complex skeletal structure (middle),
  our framework can
  (a) synthesize a high-quality novel motion, within a fraction of a second;
  (b) complete a partial motion (lower-body motion) with example motion patches;
  (c) synthesize a coherent sequence guided by a sparse set of keyframes (in blue clothes);
  (d) generate an infinitely looping animation that starts and ends with a specified pose (in blue clothes).
  }
  \Description{This is the teaser figure for the article.}
  \label{fig:teaser}
\end{teaserfigure}

\setcopyright{acmlicensed}
\acmJournal{TOG}
\acmYear{2023} \acmVolume{42} \acmNumber{4} \acmArticle{1} \acmMonth{8} \acmPrice{15.00}\acmDOI{10.1145/3592395}

\maketitle

\section{Introduction}

The generation of natural, varied, and detailed motions is a core problem in computer animation.
Acquiring large volumes of motion data via a motion capture (mocap) system or manually authoring sophisticated animations is known to be costly and tedious.
As such, motion datasets are generally limited, especially in terms of the diversity of style, skeletal structures, or creature types, which hamper the effectiveness of existing data-driven motion synthesis methods.
Advancing generative abilities of synthesizing diverse and extensive motions from limited example sequences has therefore become an important research problem.

In recent years,
deep learning 
has taken the field of computer animation by storm.
Deep learning methods have demonstrated the ability to synthesize diverse and natural motions when 
training on large and comprehensive datasets~\cite{holden2016deep, pfnn, moglow, motionclip,raab2022modi, tseng2022edge}.
More encouragingly, the success was recently reproduced in an extremely reduced setting~\cite{ganimator},
where only one sequence is provided for training, yet, the neural network learns the sample's internal distribution,
and demonstrates the ability to synthesize diverse variants of the input example sequence.
Nevertheless,
neural motion synthesis methods carry several drawbacks that limit their applicability in practice:
(i) they require long training time;
(ii) they are prone to visual artifacts such as jittering or over-smoothing;
(iii) they do not scale well to large and complex skeleton structures.

In this paper, we explore an alternative approach to the problem.
We revisit the classical idea in computer animation \--- \emph{motion nearest neighbors}~\cite{motionfield},
which dates back long before the deep learning era and 
on which the state-of-the-art industrial solution for character animation \--- \emph{motion matching} \--- was founded~\cite{motionmatching}, delivering exceptionally high-quality motion synthesis.
Motion matching produces character animations that appear natural and respond 
to varying local contexts.
Using a large mocap database as a local approximate of the entire natural motion space,
motion matching simply searches for a motion patch that best fits a given local context. 
The dependence on a large dataset is, however, at odds with our goal: we are after a generative model that ``mines'' as many \emph{diverse} motions as possible from a \emph{single} or \emph{few} examples.
Inspired by the work of \citet{dropthegan} in image synthesis,
we take the following insights for casting motion matching into our generative model and yield \emph{generative motion matching} (\name, pronounced "gem").
First,
to retain the motion quality of motion matching and inject generative capabilities,
we exploit bidirectional similarity introduced in~\cite{simakov2008summarizing}
as a new generative cost function for motion matching.
The bidirectional similarity serves the purpose of comparing the patch distribution between the example and the synthesized sequence.
Specifically,
it encourages the synthesized sequence to contain only motion patches from
the example sequences, 
and vice versa,
the examples should only contain motion patches from the synthesis.
Consequently, no artifacts are introduced in the synthesis, 
and importantly, 
no critical motion patches are lost either.
Second,
we use a multi-stage framework to progressively synthesize a motion sequence that has minimal patch distribution discrepancy with the example,
capturing patch distributions from varying temporal resolutions.
Lastly,
we utilize the observation that the generative diversity of GAN-based methods stems primarily from the unconditional noise input~\cite{dropthegan}:
we input noise to the coarsest synthesis stage, and achieve highly diverse synthesis results.

We demonstrate that \name is more than competent in producing diverse motions from only a small set of input examples.
Notably, 
compared to existing works,
\name offers several advantages:
\begin{itemize}
\item 
\name runs very fast, without any pre-training.
A motion sequence can be synthesized within a fraction of a second.

\item 
\name inherits the appealing nature of motion matching, producing motions of high quality and fidelity.

\item
\name scales smoothly to highly complex skeletons (see the character with 433 joints in Figure~\ref{fig:teaser}), where neural networks struggle~\cite{ganimator}.

\item
It is easy to extend \name to inputs with multiple sequences and encourage the synthesis to cover all examples, which is non-trivial for GAN-based methods~\cite{ganimator}.

\end{itemize}

In addition to diverse motion generation,
we also demonstrate the versatility of our generative framework by extending to an array of scenarios that are unachievable with motion matching alone, such as  
motion completion, key frame-guided generation, infinite looping, and motion reassembly, 
all enabled by the shared foundation of generative motion matching.

\section{Related work}

\begin{figure*}[t!]
  \centering
  \includegraphics[width=\linewidth]{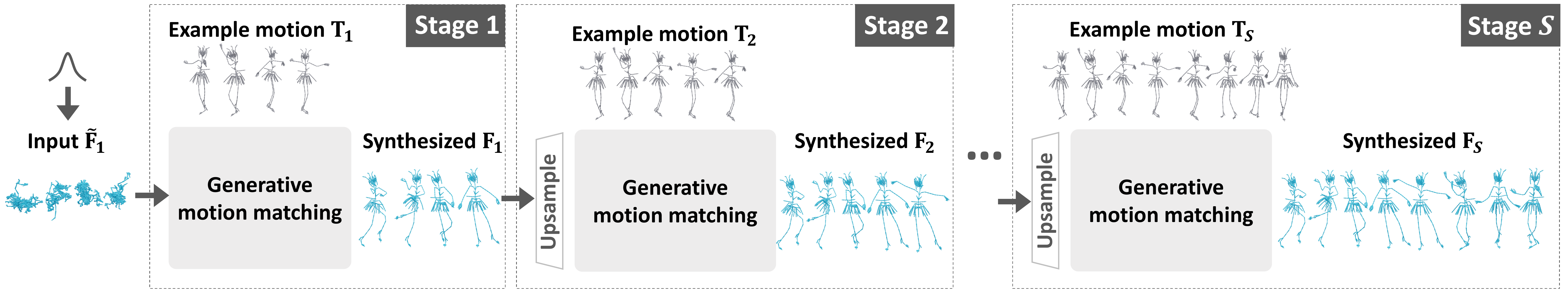}
  \caption{
  Multi-stage motion synthesis.
  Starting from the coarsest stage,
  the generative motion matching at each stage $\idstage$ takes in an upsampled version of the output from the preceding stage as the initial guess,
  refines it with motion patches in the example motion $\motionfeatin_\idstage$,
  and outputs a finer motion sequence $\motionfeatout_\idstage$.
  Note the coarsest stage is purely generative, as the input is merely a Gaussian noise. 
  }
  \label{fig:overview}
\end{figure*}

We review the most related work on kinematics-based motion synthesis.
We also briefly cover recent advancements in image synthesis, particularly patch-based ones, from which we take inspiration.

\paragraph{Motion Synthesis}
Generating novel motions via diversifying existing ones can date back to the work of ~\citet{perlin1996improv}, where the Perlin noise~\cite{perlin1985image} is added to motion clips for obtaining variants with local diversity. 
\citet{pullen2002motion} show that mocap data can be used to enhance a coarse key-framed motion, by matching low-frequency patches and blending high-frequency details. 
\citet{motiontexture} 
construct a graph model by matching similar patches in the dataset,
and create a stochastic model for generating random samples with local and structural variations.
Due to the use of a linear dynamics model, the model requires a large training dataset to achieve satisfactory results and faces a dilemma of quality and diversity.
Contemporarily, motion graphs~\cite{kovar2002motion, lee2002interactive, arikan2002interactive} use a similar discrete graph model while keeping it deterministic, namely a state machine, and demonstrate characters that respond interactively to the user input.
However, the discrete space inherently limits their agility and responsiveness. 
Hence, efforts on summarizing large datasets into statistical models have also been made since then ~\cite{pullen2000animating, brand2000style, bowden2000learning, grochow2004style, chai2007constraint, wang2007gaussian}. 
Instead of sorting the dataset into an organized but discrete structure or a statistical model, 
motion nearest neighbors~\cite{motionfield, levine2012continuous} operate directly on contiguous \emph{motion fields} to learn a control policy that interpolates the nearest neighbors of the current pose.
Following that,
~\cite{motionmatching} introduce Motion Matching, which is a method searching a large database of animations for the animation which best fits the given context. 
This method has quickly been
adopted by many studios due to its simplicity, flexibility, controllability, and the quality of the motion it produces~\cite{harrower2018real, buttner2019machine}.
Motion matching plays back the animation data
stored in the database as-is,
rendering it the de facto state-of-the-art in the industry.
Nevertheless, its goal differs significantly from ours, as we target a generative model that synthesizes diverse motions from examples.

Recent advancements in deep learning also greatly impact the motion synthesis field. 
Early attempts
\cite{holden2015learning,holden2016deep} use deep neural networks to learn from animation data. 
Deep neural networks can learn a strong prior from a large dataset~{~\cite{humor, nemf}, solving many ill-posed generative tasks including motion prediction~\cite{fragkiadaki2015recurrent, pavllo2018quaternet}, motion in-betweening~\cite{harvey2020robust, duan2022unified, qin2022motion}, motion reassembling~\cite{motionpuzzle, Chimera}, text-guided motion synthesis~\cite{motionclip, mdm, zhao2023modiff}, etc. 
\citet{holden2015learning,holden2016deep} apply modern deep learning techniques for learning from animation data. 
Meanwhile, combining motion matching with deep learning has also resulted in variants that are computationally less expensive~\cite{holden2020learned} and more versatile~\cite{habibie2022motion}. 
Notably,
all these works require a large and comprehensive dataset for training.
Another noticeable line of work adapts deep reinforcement learning techniques to train a physically simulated character with a small set of example motions~\cite{deepmimic, amp}. 
More recently, \cite{ganimator}, the most related work to us, proposes to use a patch GAN-based~\cite{pix2pix2017, singan} approach to train a generative model with a single example. 
Concurrently,
\citet{raab2023single} introduce a diffusion-based model that learns the internal motifs of a single motion clip for producing diverse outputs.
Nonetheless,
these two methods struggle to produce results with sharp motions, and are not suitable for training with multiple examples due to the discontinuous underlining 
 latent space presented to it.
We show the superiority of our method over GANimator by an in-depth comparison in \Cref{sec:comparisons}.

\paragraph{Image Synthesis} 
Our work adopts several algorithmic designs from texture image synthesis, that shares a similar goal with motion synthesis. 
For an in-depth survey of this extensive body we refer readers to surveys \cite{wei2009state, barnes2017survey}.
The image pyramid, also known as progressive generation~\cite{karras2018progressive} in deep learning, had been used in texture synthesis long ago. 
\citet{heeger1995pyramid, de1997multiresolution} use a Laplacian pyramid~\cite{burt1987laplacian} for texture synthesis, realizing progressive generation on the spatial frequency domain. \citet{wei2000fast} use a Gaussian pyramid for a similar purpose. \citet{multiscale} push the multi-scale generation to a new height,
where gigapixel-sized images with great details can be synthesized.
We also adopt the progressive synthesis, allowing our generative motion matching module to capture details of different levels. 
Progressive synthesis has also become popular in today's era of deep learning,
leading to impressive generative models that learn to progressively refine random noise into images resembling a single natural image. 
Specifically,
a series of GANs~\cite{goodfellow2014generative} are trained to capture the patch distribution of the example at varying scales.
Following that, \citet{dropthegan} show that
bidirectional visual similarity~\cite{simakov2008summarizing} can serve the purpose of measuring the patch distribution discrepancy between the example and the synthesized image,
leading to diverse images of much higher quality and fast synthesis, compared to GAN-based methods.

\section{Method}

We elaborate details of our generative framework,
that can synthesize high-quality motions resembling given examples, in large quantities and varieties.
Although our method can take as input \emph{multiple} exemplar motions,
in our coverage,
to ease the understanding of the algorithm,
we mainly describe in the \emph{single} input setting.
Moreover, the synthesized motion does not need to match exactly the length of the example and can be of arbitrary length.

\subsection{Motion Representation}
A motion sequence is defined by a temporal set of $\nbframein$ poses
that each consists of root joint displacements $\rootpos \in \realnum^{\nbframein \times 3}$ and joint rotations $\jointrot \in \realnum^{\nbframein \times \nbjoint \rotdim}$,
where $\nbjoint$ is the number of joints and $\rotdim$ is the number of rotation features.
Instead of directly using the global root displacements $\rootpos$, 
we convert $\rootpos$ to local root displacements $\rootdelta$,
that are temporal-invariant and calculated as the difference between every two consecutive poses.
The joint rotations are defined in the coordinate frame of their parent in the kinematic chain,
and we use the 6D rotation representation (i.e., $\rotdim = 6$) proposed by \citet{zhou6d}.

As human eyes are rather sensitive to implausible interactions between the end-effector and the ground,
existing neural-based methods usually establish geometric losses on the locations and velocities of the end-effectors, i.e., the foot contact loss~\cite{shi2020motionet, ganimator, mdm},
whereas 
our method does not demand such a design
as the internal structure of exemplar motion patches,
such as the high correlation between the end-effectors and the root motion,
are inherently preserved in the synthesis.
That said,
it is also trivial to incorporate foot contact labels as in~\cite{ganimator} into our representation for improvements in \emph{rare} cases,
where sliding feet could occur in the synthesis with large root-motion examples.
The contact label enables the IK post-process to avoid floating feet.
Specifically,
the contact labels $\contactlabel$ can be easily retrieved from the input motion by setting a threshold of the magnitude of the velocity. 
Assume the number of foot joints is $\ncontact$, for each foot joint at a timestamp, we calculate a binary vector and append it to the per-frame feature (See Figure~\ref{fig:patch_extraction} and~\ref{fig:match_and_blend}).

For convenience,
we let
$\motionmetric_{\nbframe} \equiv \realnum^{\nbframe \times \left( \nbjoint \rotdim + 3 + \ncontact \right)}$ 
denote the metric space of concatenated motion features of $H$ frames, 
$\motionfeatin \equiv \left[ \jointrot, \rootdelta \right] \in \motionmetric_{\nbframein}$ the original input motion features,
$\motionfeatin_i \in \motionmetric_{\nbframein_i}$ a corresponding downsampled version of the input,
$\motionfeatout \in \motionmetric_{\nbframeout}$ the synthesized motion features of $\nbframeout$ frames,
and 
$\motionfeatout_\idstage$ a corresponding downsampled version of $\motionfeatout$.

\subsection{Multi-stage Motion Synthesis}
Figure~\ref{fig:overview} presents the overall pipeline of our approach,
which consists of $\nbstage$ stages to progressively synthesize a motion of $\nbframeout$ frames.
Specifically,
given an input motion,
we build an exemplar pyramid $\{ \motionfeatin_1, ...,  \motionfeatin_\nbstage \}$,
where $\motionfeatin_\nbstage = \motionfeatin$ is the original input sequence and $\motionfeatin_{\idstage} \in \motionmetric_{\nbframein_\idstage}$ is $\motionfeatin_{\idstage+1}$ downsampled by a factor $\downsamplef > 1$.
Then,
a synthesis pyramid $\{ \motionfeatout_1, ..., \motionfeatout_\nbstage \}$,
where $\motionfeatout_\nbstage \in \motionmetric_\nbframeout$ is the final synthesized sequence of $\nbframeout$ frames and $\motionfeatout_{\idstage} \in \motionmetric_{\nbframeout_\idstage}$ is an intermediate sequence of $\nbframeout \cdot \downsamplef^{\idstage - \nbstage}$ frames,
will be synthesized in a coarse-to-fine manner, starting from the coarsest stage and up to the finest.
At each stage $\idstage$,
the generative motion matching module (Section ~\ref{sec:patch_generation}) takes in an upsampled version of the output from the preceding stage as the initial guess, $\tilde{\motionfeatout}_{\idstage} = \motionfeatout_{\idstage-1} \upsample$,
refines it with exemplar motion patches in $\motionfeatin_\idstage$,
and outputs a finer motion sequence $\motionfeatout_\idstage$.
Note that
the synthesis
at the coarsest stage is purely generative,
as
the input is merely a noise drawn from a Gaussian distribution,
i.e.,
$\tilde{\motionfeatout}_1 \sim \mathcal{N}(\mu,\,\sigma^{2}) \in \motionmetric_{\nbframeout_1}$.

\begin{figure}[t!]
  \centering
  \includegraphics[width=\linewidth]{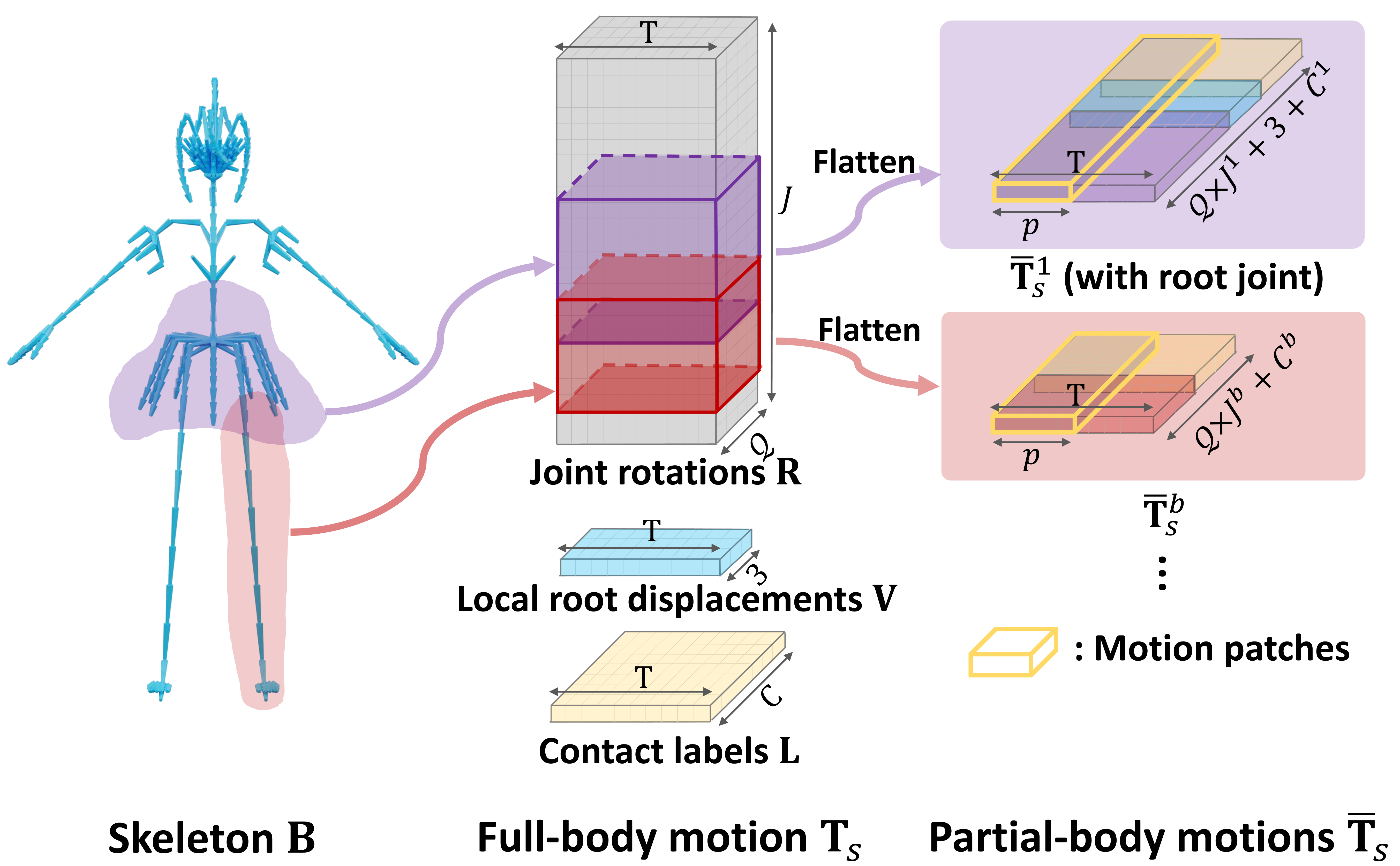}
  \caption{
  Skeleton-aware motion patch extraction.
  The skeleton is partitioned into several overlapping skeletal parts (two coloreds on the left),
  with which the full-body motion can be split into a set of partial-body motions accordingly.
  Then motion patches (yellow boxes) with a temporal size of $\patchsize$ frames can be extracted from each partial-body motion.
  }
  \label{fig:patch_extraction}
\end{figure}

\subsection{Generative Motion Matching}
\label{sec:patch_generation}
Typically,
patch-based image synthesis consists of three steps, namely the patch extraction, nearest neighbor matching, and blending,
that work in sequence to produce a converging result in multiple iterations.
Our method follows a similar approach but with algorithmic designs specific to our task.
Specifically,
at each stage $\idstage$,
the following steps are invoked sequentially during $\nbiter$ iterations.

\paragraph{Skeleton-aware Motion Patch Extraction.}
A motion patch can be
defined trivially as a sub-sequence of $\patchsize$ consecutive frames in the example sequence,
which is a common practice in motion synthesis~\cite{ganimator, motionmatching}.
While our approach can simply work with this definition,
we further propose to extract skeleton-aware motion patches from the motion sequence,
which decomposes the skeleton into multiple sub-sets, 
i.e., skeletal parts,
instead of treating it as a whole,
and eventually leads to more \emph{diverse poses}.
Specifically,
let $\skeleton$ denote the skeletal tree used by the example full-body motion $\motionfeatin_\idstage$,
a set of skeletal parts 
$\{ \skeletalpart_1, ..., \skeletalpart_\nbpart \}$,
where $\skeletalpart_\idpart \subset \skeleton$ is a sub-tree of the whole skeleton and has $\nbjoint^{\idpart}$ joints,
can be defined to divide the full-body motion into a set of partial-body motions $\{ \partialmotion^1_\idstage, ..., \partialmotion^\nbpart_\idstage \}$,
from which we crop sub-sequences of $\patchsize$ frames with stride size $1$ as our motion patches (See Figure~\ref{fig:patch_extraction}).

\begin{figure}[t!]
  \centering
  \includegraphics[width=\linewidth]{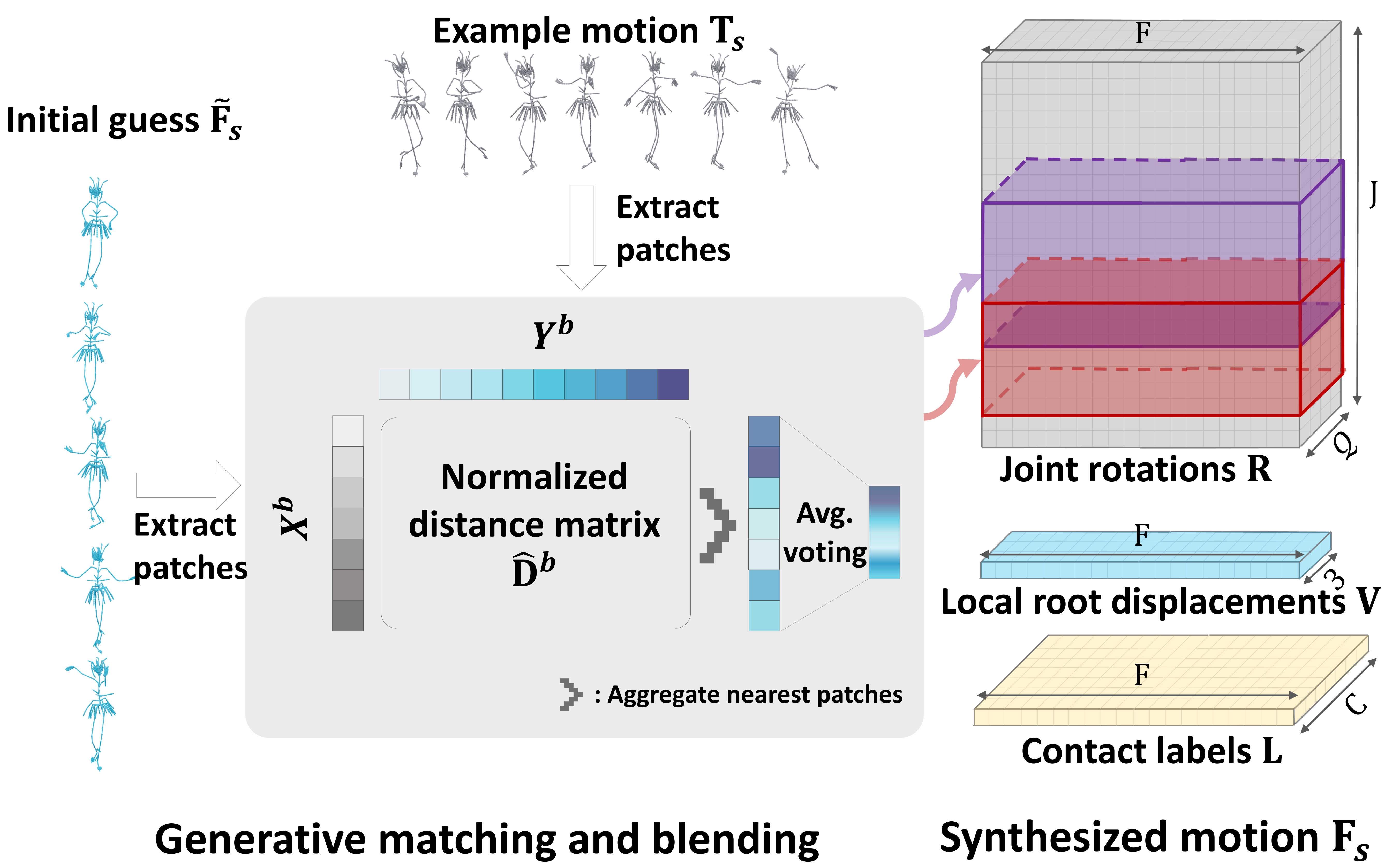}
  \caption{
  Generative matching and blending. 
  Each motion patch in the initial guess 
  finds the best-matched motion patch in the example motion,
  according to the normalized distance matrix.
  Then we blend the overlapping matched patches to form a novel partial motion. 
  Finally, we blend multiple resultant partial motions to get the final full-body motion (see right). 
  }
  \label{fig:match_and_blend}
\end{figure}

Usually,
skeletons across different animations do not necessarily follow specific rules and can be of extremely high variability,
for example,
bipeds vs. hexapods, 
a pure biological skeleton vs. one with more artistic joints, etc.
Hence,
our approach allows the user to manually divide the whole skeletal structure into sub-parts with overlapping joints, similar as in~\cite{motionpuzzle, Chimera}.

\begin{figure*}[th!]
  \centering
  \includegraphics[width=\linewidth]{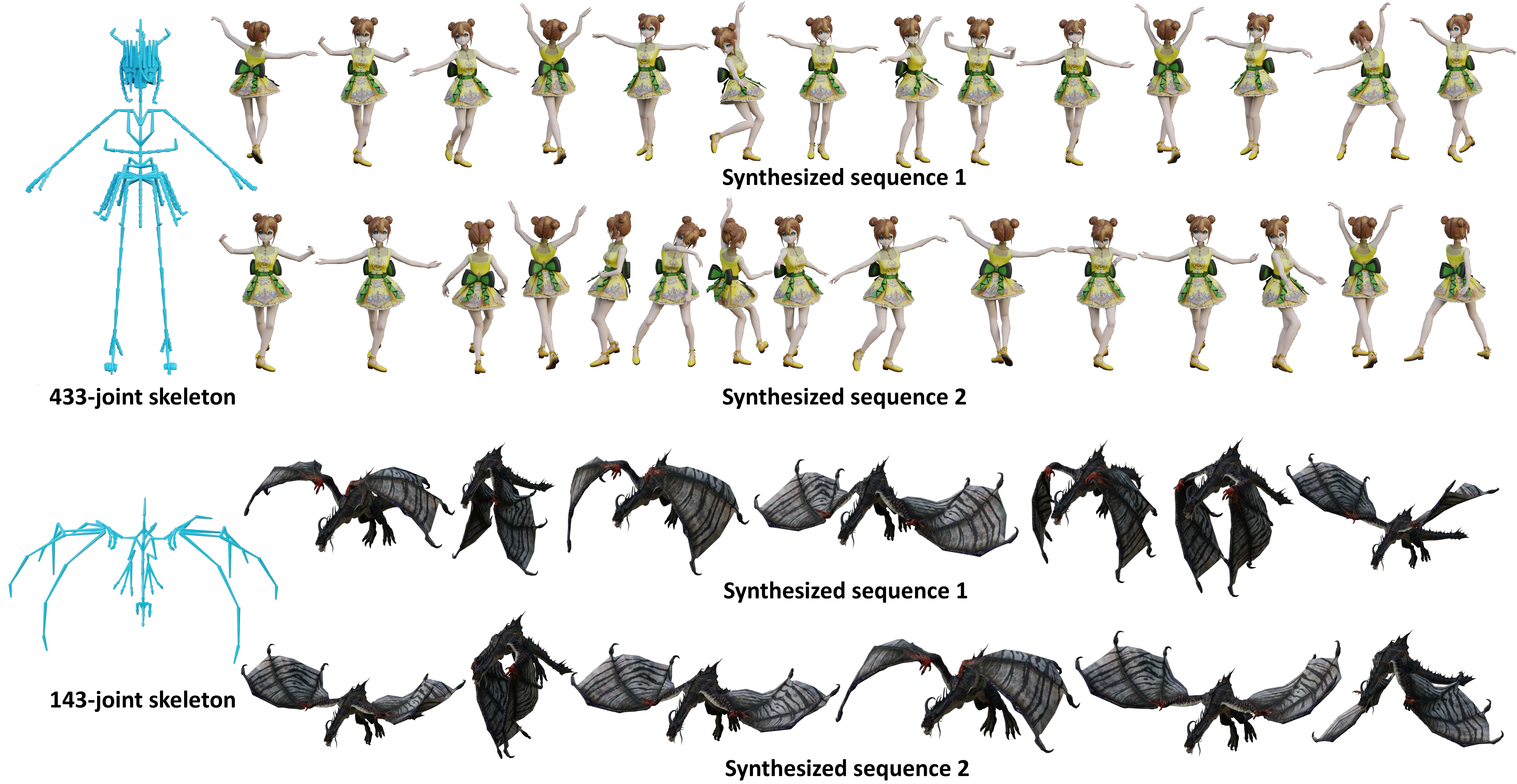}
  \caption{
  From a single example,
  our framework generates, within a second, diverse motion sequences for even highly complex and large skeletons, including the animation of the clothes and the wings.
  Please refer to the accompanying video for more animation results. 
  }
  \label{fig:randomgenetaion}
\end{figure*}

\paragraph{Generative  Matching.}
Let $\qset$ denote the set of motion patches extracted from $\tilde{\motionfeatout}_{\idstage}$,
$\kset$ the set of motion patches extracted from the example motion $\motionfeatin_{\idstage}$.
We calculate pairwise patch distance matrices using squared-$L2$ distance,
which provides the foundation for the measurement of the similarity between each exemplar motion patch and each synthesized motion patch.
Note that the patch distance matrix is calculated \emph{per} skeletal part:
\begin{equation}
\begin{gathered}
   \dist_{i,j}^{b} = \| \qset_{i}^{b} - \kset_{j}^{b} \|^2_2,
   \label{eq:patch_dist_matrix}
\end{gathered} 
\end{equation}
where $\qset^{\idpart}$ and $\kset^{\idpart}$ denote the set of motion patches extracted from corresponding partial-body motions.
Then, the bidirectional similarity as in \cite{simakov2008summarizing, dropthegan} is introduced to encourage
that all exemplar motion patches appear in the synthesis and all motion patches in the synthesis do not deviate from the example,
i.e., high completeness and coherence.
This is achieved by normalizing the distance matrices using a per-example-patch factor:

\begin{equation}
\begin{gathered}
   \simi_{i,j}^{b} =  \frac{\dist_{i,j}^{b}}{(\alpha + \min_{\ell}(\dist_{\ell,j}^{b}))} ,
\label{eq:distance_metric}
\end{gathered} 
\end{equation}
where $\alpha$ controls the degree of completeness,
and smaller $\alpha$ encourages completeness.
An in-depth study on the effect of $\alpha$ is conducted in \Cref{hyperparameters}.

\paragraph{Blending.}
For each motion patch in $\qset^b$, we find its nearest (as defined by Equation~\ref{eq:distance_metric}) motion patch in $\kset^b$, 
and then blend the values of collected motion patches using average voting,
forming a synthesized partial-body motion $\bar{\motionfeatout}^b$.
Finally,
we average the values over overlapping joints between skeletal parts
to assemble all synthesized partial-body motions into the final result $\motionfeatout_\idstage$ (See Figure~\ref{fig:match_and_blend}).

\subsection{Extension to More Settings}
\label{sec:more_setting}

Our method can also be easily extended for various settings.

\paragraph{Skeleton Partition for Motion Patches.}
In addition to the skeleton-aware motion patch defined above,
our method can also work with the traditional definition of a motion patch, i.e., treating the skeleton as a whole and then extracting $\patchsize$ consecutive poses.

\paragraph{Multiple Examples.}
As aforementioned,
our method can not only be applied to a single motion input but also works with multiple sequences of different numbers of frames.
This can be achieved by simply extracting motion patches from all input motions to form the set of exemplar patches used in Equation~\ref{eq:patch_dist_matrix}.
In this setting, the completeness control knob $\alpha$ plays a crucial role in ensuring that all motion patches across examples are utilized in the output.

\paragraph{Heterogeneous Skeletons.}
Interestingly,
under the setting of multiple examples,
the skeletons across different motions do not necessarily share the same one.
For example, 
we can take motion clips of a monster and a zombie, 
and synthesizes a moving Frankenstein, 
via harmonizing different partial-body motions extracted from these two creatures.
To achieve this, the user can split the skeleton of each input with overlap and manually specify the skeletal parts of interest to be used in the generative motion matching and blending. 
The overlapping region plays a crucial role in bridging different skeletal parts during the generation process. 
Then, our method can synthesize the novel motion for the new creature by combining the partial-body motions as discussed in Section~\ref{sec:patch_generation}.

\section{Experiments}
\label{experiments}

\begin{figure*}[]
  \centering
  \includegraphics[width=\linewidth]{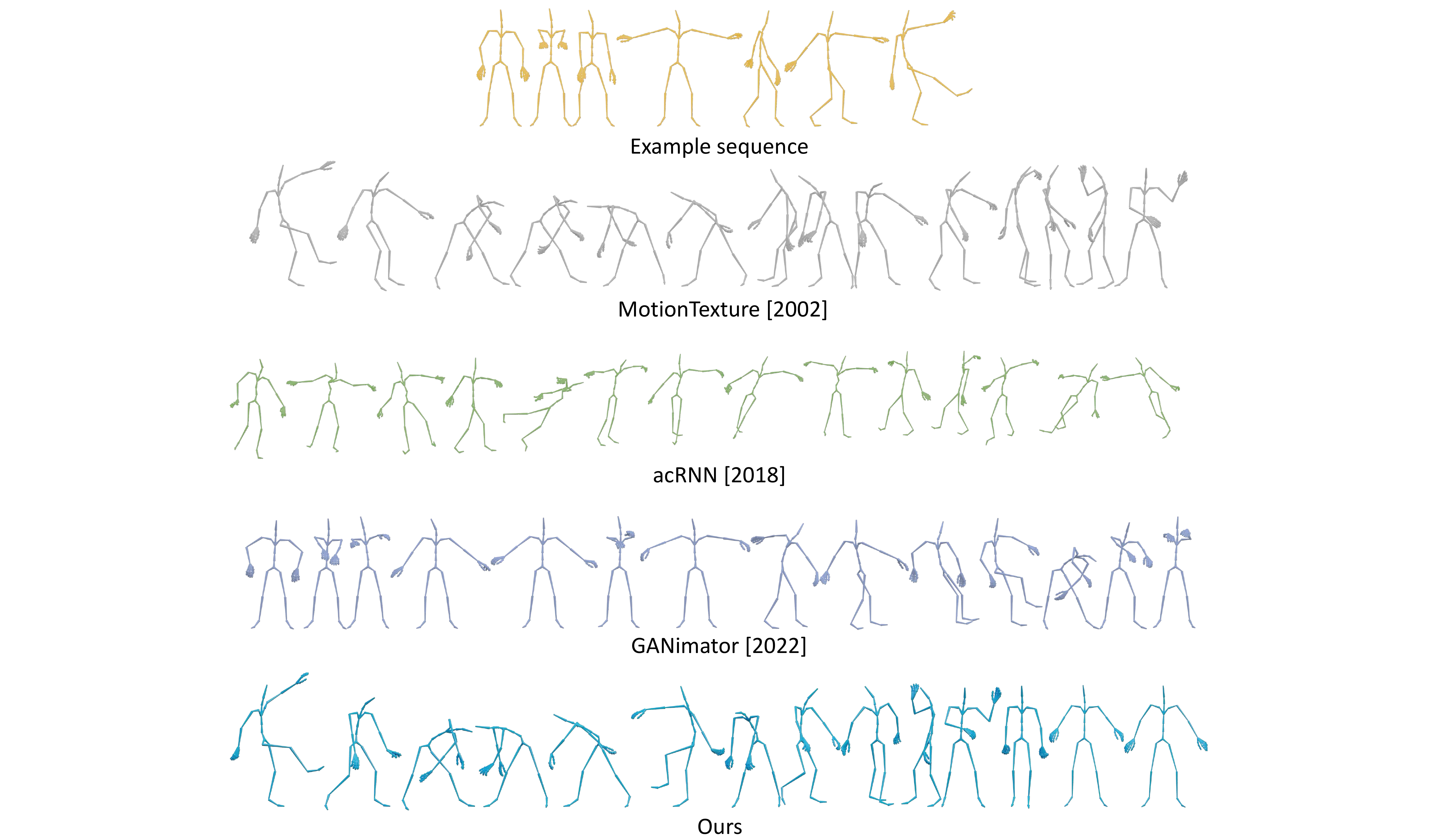}
  \caption{
  Visual comparisons. 
MotionTexture~\shortcite{motiontexture} generates motions with unnatural transitions. acRNN~\shortcite{acrnn} produces noisy motions or sometimes quickly converges to a static pose. GANimator~\shortcite{ganimator} struggles to handle complex skeletons, producing over-smoothing results. 
Our method outperforms these methods with diverse and high-quality results, where highly dynamic motions are well preserved. 
Please refer to the accompanying video for animation results.
  }
  \label{fig:comparisons}
\end{figure*}

We evaluate the effectiveness of our method on example-based motion synthesis, compare to other motion generation
techniques,
and demonstrate its versatility by applying it in various settings and applications. We highly recommend readers refer to the accompanying video for more qualitative evaluations.
\emph{Code and data will be released
to ease the understanding of our implementation and facilitate future studies.}

\paragraph{Data.}
We collected a diverse set of example animations featuring varied motion styles and highly complex and large skeletal structures from Mixamo~\shortcite{mixamo} and Truebones~\shortcite{truebones}.
The motion styles we experimented
with include sharp motions of a popping dance, subtle motions of fanning wings, etc.
Some examples are authored with highly sophisticated skeletal structures, such as the 433-joint and 143-joint skeletons as visualized in Figure~\ref{fig:randomgenetaion}.
The number of frames ranges from 140 to 1000 frames at 30 fps.

\paragraph{Implementation Details.}
Our framework is lightweight and does not require any training.
Due to its simplicity and efficiency,
we simply implement our method with Python.
We also develop an add-on in the open source software Blender~\cite{blender}, which is ready to take animations from users and synthesizes diverse and high-quality variants.
In our implementation,
a motion sample with around 1000 frames can be generated in $\sim0.2$s with an Apple M1 CPU or $\sim0.05$s with a modern GPU (NVIDIA V100).
By default, we run experiments using an Apple M1 CPU,
except that the comparison experiments are conducted using an NVIDIA V100 GPU for fair comparisons with neural network-based methods. 
We set the length of $\motionfeatin_1$ at the coarsest stage to $K$ times the patch size $\patchsize$. 
Thus the receptive field (a similar concept as in the image) always occupies the same proportion of example motions with different lengths.
Then, $\motionfeatin_1$ is gradually upsampled using the factor $\downsamplef$
until it reaches the final length of $\motionfeatin_S$.
Unless otherwise specified,
we use a  patch size $\patchsize=11$, $K=4$, a completeness control knob $\alpha=0.01$ and a number of iterations $\nbiter = 5$. These are the empirically best hyper-parameters we have found. For more discussions on the hyper-parameters, please refer to the supplementary material.

\begin{figure*}[]
  \centering
  \includegraphics[width=\linewidth]{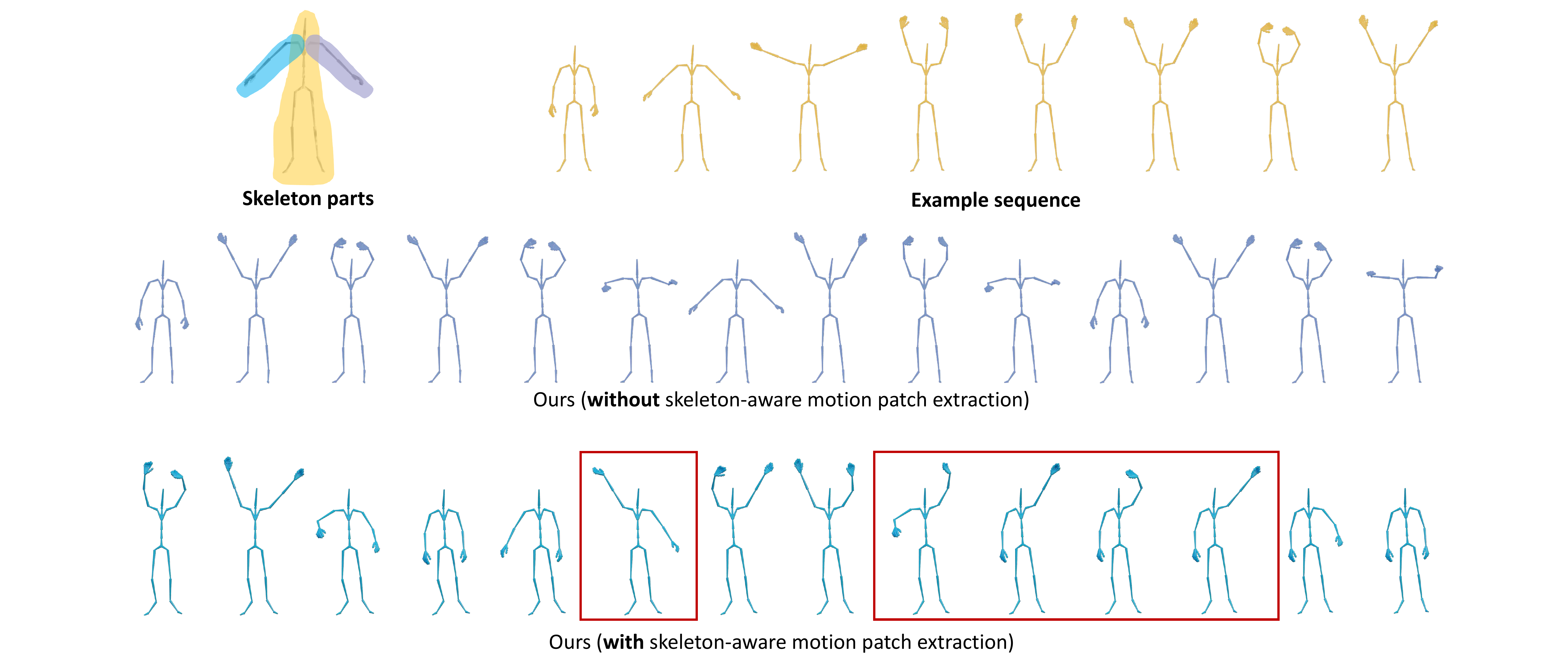}
  \caption{The effectiveness of skeleton-aware motion patch extraction. 
  An artist manually divides the skeleton into three overlapping parts (top left).
  Given an example of a character waving two hands simultaneously, only a sequence with two waving hands can be synthesized without the skeleton-aware component. However, with the skeleton-aware motion patch extraction, a more diverse sequence, including waving with only one hand (in red boxes), can be generated.
  }
  \label{fig:skeleton_aware}
\end{figure*}

\subsection{Novel Motion Synthesis}
\label{randomsynthesis}

We first evaluate the performance of our framework on novel motion synthesis,
and compare it to a classical statistical model and recent neural-based models, namely MotionTexture~\cite{motiontexture}, acRNN~\cite{acrnn} and GANimator~\cite{ganimator}.

\paragraph{Settings}
Although our method can use multiple inputs, 
for fair comparisons,
we conduct the evaluation on divers motion synthesis from a \emph{single} example and disable the skeleton-aware motion patch extraction.
We use three example sequences containing highly dynamic and agile movements for evaluation. The character consists of 65 joints, and each sequence has around 500 frames.
For each example sequence,
we use all methods to synthesize a novel sequence that doubles the length of the example.

\paragraph{Qualitative Comparison}
\label{sec:comparisons}
(i)
To accomplish diverse motion synthesis, \emph{MotionTexture} organizes similar motion patches from training motions into linear dynamics models known as textons, and models the probability of transition between textons using a Markov Chain. 
However, it faces the challenge of balancing diversity and quality, particularly when there is only one example sequence due to the choice of linear dynamics model.
We follow the procedure as done in~\cite{ganimator} to apply MotinTexture to a single example.
As a result, MotionTexture produces unnatural transitions between textons.
(ii)
\emph{acRNN} uses an RNN-based network structure.
The lack of data leads to a model that is prone to overfitting and is not robust to perturbation or error accumulation.
Consequently, acRNN can only stably generate a limited number of frames.
(iii)
\emph{GANimator} utilizes a series of GANs to capture the distribution of motion patches at different scales, in order to progressively synthesize motions that closely resemble the input.
In our experiments with complex and large skeletons, as shown in \Cref{tab:differentskeletons}, GANimator struggles to produce high-quality results, often resulting in jittery or over-smoothed motions. Additionally, it requires a significant amount of training time, typically from several hours to a day.
In contrast,
our method can adapt to these complex skeletal structures and various motion styles, and synthesize diverse and high-quality variations as shown ~\Cref{fig:comparisons}.
Notably, highly dynamic motions, in particular sharp and agile movements, are well preserved in our synthesized results. 
For more qualitative results, please refer to the accompanying video.

\begin{figure*}[t!]
  \centering
  \includegraphics[width=0.95\linewidth]{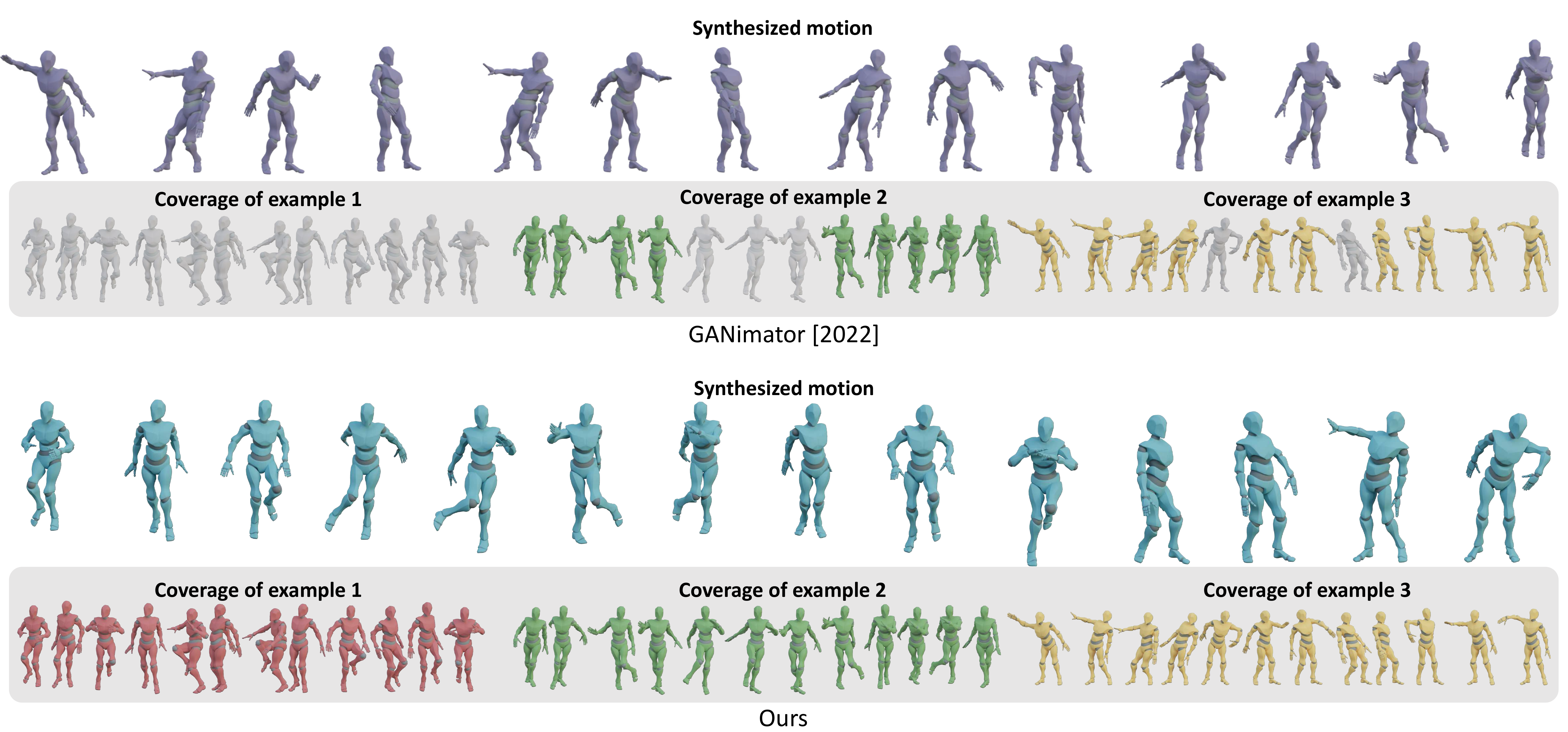}
  \caption{
    Comparison under the multi-example setting.
    GANimator struggles to incorporate all examples, resulting in the loss of a significant portion of exemplar motion patches in the synthesis (marked in gray) and a low coverage score. In contrast, our method effectively covers all examples (marked as colorized), resulting in high coverage score. }
    \label{fig:coverage}
\end{figure*}

\begin{table}[t!]
\caption{
Quantitative comparisons on single example-based generation.
}
\footnotesize
\setlength{\tabcolsep}{1.6pt}
\centering

\begin{tabular}{lcccccc} 
\toprule
  & Coverage & \begin{tabular}[c]{@{}c@{}}Set \\Div.\end{tabular} & \begin{tabular}[c]{@{}c@{}}Global \\Patch Dist.\end{tabular} & \begin{tabular}[c]{@{}c@{}}Local \\Patch Dist.\end{tabular} & \begin{tabular}[c]{@{}c@{}}Training\\Time \end{tabular} & \begin{tabular}[c]{@{}c@{}}Inference\\Time \end{tabular} \\ 
\midrule
MotionTexture~\shortcite{motiontexture} & 84.12 & 0.05 & 1.12 & 1.13 & 32.3s & 0.03s \\
MotionTexture (Single) & 100.00 & 0.01 & 0.45 & 0.44 & 0.08s & 0.07s \\
acRNN~\shortcite{acrnn} & 5.13 & 0.75 & 13.62 & 13.55 & 25 hrs & 0.21s\\
GANimator & 49.07 & 0.24 & 2.18 & 2.08 & 6 hrs & 0.12s\\
Ours & 99.89 & 0.28 & 0.22 & 0.18 & N/A & 0.08s\\ 
\bottomrule
\end{tabular}

\label{tab:single_motion_gen}
\end{table}%

\paragraph{Quantitative Comparison} Measuring the quality of generated results against a few examples is known to be difficult~\cite{ganimator}. 
As one of the pioneers, GANimator uses a combination of established metrics, namely coverage, diversity, and reconstruction loss, to rate the performance, since a single metric does not suffice the need of measuring the overall quality. 
However, the reconstruction loss is only suitable as a quality indicator for neural network-based methods. 
The diversity is measured with the average distance between generated motion patches and their nearest neighbor in the examples, and different motion patch size corresponds to local and global diversity. 
As a result, they tend to increase if the generated motion becomes unnatural, and favor results with minor perturbation or over-smoothed results generated by neural networks. 
We thus use neutral names for the diversity metrics, namely \emph{local patch distance} and \emph{global patch distance} in our experiments. 
We refer readers to \cite{ganimator} for more details about these metrics.
In addition, 
following the well-established metric in 2D image synthesis~\cite{singan}, %
we also report the \emph{set diversity}, which measures the diversity among the generated results and is calculated as the averaged standard deviation of the rotation of each joint over 200 synthesized motions and normalized by the standard deviation of all joint rotations of the input example.
Note that, while this metric also has a preference for noisy output, we mainly rate the methods using the combination of the coverage and set diversity.

The quantitative comparison results are shown in Table~\ref{tab:single_motion_gen}.
Notably,
our method produces a significantly high coverage score, while still exhibiting sufficiently diverse results (evidenced by a high set diversity score). For a more comprehensive comparison of the quality, we refer the readers to the accompanying video.
Note, we also report the computation time in \Cref{tab:single_motion_gen}, where we can see that our method is highly efficient, as it is both training-free and extremely fast during inference.

\subsection{More Generation Settings}
\label{more_experiments}
In addition to the basic setting used above,
we further evaluate our method in the following aspects.

\paragraph{Skeleton-aware Motion Patches.}

In addition to the temporal axis, our method can also extract motion patches from examples along the skeletal axis, thus allowing obtaining diversity also on the spatial dimension as shown in \Cref{fig:skeleton_aware} and the accompanying video.

\paragraph{Multiple Examples.}

Unlike existing methods, 
which struggle when presented with multiple example sequences due to the lack of explicit encouragement of completeness,
our method can handle multiple examples with the completeness control knob in Equation~\ref{eq:distance_metric}.
We collect five dancing sequences ranging from 120-220 frames at 30 fps. It can be seen in \Cref{tab:multimotiongencoverage} that when more examples are given, existing methods generate results with lower coverage while our method remains a high coverage.
MotionTexture produces unnatural transitions, similar to its results in the single-example setting.
acRNN fails on the task and produces noisy motions due to the diverse but scarce motion data.
GANimator requires a corresponding pre-defined latent variable for each sequence. However, the structure of the given sequences is not taken into consideration for defining these latent variables, hindering the network from generating various and complete samples.
In contrast,
our method 
produces motions that cover a large portion of the examples with a properly set completeness control knob as shown in \Cref{fig:coverage} and the accompanying video.

\begin{table}[t!]

\caption{
Coverage rates of different numbers of example sequences.
}
\footnotesize
\setlength{\tabcolsep}{3pt}
\centering

\begin{tabular}{lcccc} 
\toprule
 & \multicolumn{4}{c}{Number of Example Sequences} \\ 
\cline{2-5}
 & 2 & 3 & 4 & 5 \\ 
\midrule
MotionTexture~\shortcite{motiontexture} & 100 & 29.19 & 10.03 & 27.69 \\
acRNN~\shortcite{acrnn} & 7.34 & 4.02 & 1.38 & 0.41 \\
GANimator~\shortcite{ganimator} & 63.55 & 23.90 & 17.30 & 16.22 \\
Ours & 99.71 & 99.95 & 99.91 & 99.64 \\
\bottomrule
\end{tabular}

\label{tab:multimotiongencoverage}
\end{table}%

\begin{table}[t!]
\caption{
Coverage rate on skeletons with different complexity.
}
\footnotesize
\setlength{\tabcolsep}{8pt}
\centering

\begin{tabular}{lccc} 
\toprule
 & \multicolumn{3}{c}{Number of Joints} \\ 
\cline{2-4}
 &  24 & 65 & 433 \\ 
\midrule
GANimator~\shortcite{ganimator} & 92.10 & 44.10 & 2.38 \\
Ours & 97.82 & 99.84 & 86.90 \\
\bottomrule
\end{tabular}

\vspace{-8px}

\label{tab:differentskeletons}
\end{table}%

\paragraph{Complex Skeletons.}
Our method can work with skeletons of high complexities (See Figure~\ref{fig:randomgenetaion}),
on which the GAN-base method GANimator fails to produce reasonable results as demonstrated in the accompanying video.
Specifically,
we experiment with skeletons consisting of 24, 65, and 433 joints.
It can be seen in \Cref{tab:differentskeletons} that GANimator performs normally on the 24-joint skeletons, while
its performance drops dramatically when presented with complex skeletons.
Whereas our method maintains a consistent performance for different skeletons,
evidenced by the fluttering effects of the skirt and dragon wings in the accompanying video.

\begin{table}[t!]
\caption{
The different settings of hyperparameters.
}
\footnotesize
\setlength{\tabcolsep}{1.5pt}
\centering

\begin{tabular}{lccccc} 
\toprule
 & Coverage & Set Diversity & Global Patch Dist. & Local Patch Dist. \\ 
\midrule
Ours (not use $\alpha$) & 87.45 & 0.27 & 0.18 & 0.17 \\
Ours ($\alpha = 5$) & 87.89 & 0.27 & 0.18 & 0.17 \\
Ours ($\alpha = 0.5$) & 88.47 & 0.27 & 0.18 & 0.17 \\
Ours ($\alpha = 0.05$) & 93.80 & 0.27 & 0.18 & 0.17 \\
Ours ($\alpha = 0.005$) & 99.96 & 0.27 & 0.30 & 0.23 \\ 
Ours ($\alpha = 0.0$) & 36.78 & 0.25 & 2.17 & 1.60 \\ 
\midrule
Ours ($K = 20$) & 87.74 & 0.25 & 1.03 & 0.71 \\
Ours ($K = 15$) & 92.04 & 0.26 & 0.73 & 0.54 \\
Ours ($K = 10$) & 96.07 & 0.27 & 0.42 & 0.36 \\
\midrule
Ours ($\patchsize = 23$) & 99.85 & 0.26 & 0.27 & 0.24 \\
Ours ($\patchsize = 17$) & 99.88 & 0.26 & 0.23 & 0.21 \\
Ours ($\patchsize = 5$) & 99.66 & 0.27 & 0.41 & 0.32 \\
\midrule
Ours ($\downsamplef = 2$) & 99.61 & 0.27 & 0.22 & 0.20 \\
Ours ($\downsamplef = 4$) & 99.16 & 0.27 & 0.34 & 0.28 \\
Ours ($\downsamplef = 8$) & 97.97 & 0.26 & 0.57 & 0.42 \\ 
\bottomrule
\end{tabular}

\label{tab:hyperparameters}
\end{table}%

\subsection{Effects of Hyper-parameters}
\label{hyperparameters}
Our framework involves several hyper-parameters during the synthesis process. 
In this section, we discuss the effects of these hyper-parameters. The quantitative results are presented in \Cref{tab:hyperparameters}.

\paragraph{Effects of $\alpha$}
As rarely-appearing patches have a larger minimal distance,
the completeness of the synthesis is encouraged by normalizing the distance of patches extracted from examples with their minimal distance to the initial guess. 
Therefore, the $\alpha$ in Equation 2 serves as a control knob for the completeness of exemplar patches in the synthesized result.
As it restricts the lower bound of the normalizing denominator, a smaller $\alpha$ value encourages more preservation of the example content in the synthesis.
As shown in Table~\ref{tab:hyperparameters}, when $\alpha$ decreases to a certain level, a higher coverage score is achieved. However, if the value $\alpha$ is too small, an excessive emphasis on completeness (especially for patches with almost zero distance to the generated motion) can overwhelm the similarity measure used for the matching process, resulting in unstable generation and low-quality motion (evidenced by low coverage and high patch distances of the corrupted results)."

\paragraph{Effects of $K$}
The ratio of the patch size to the length of input example motion at the coarsest stage 
controls the receptive field for synthesis, similar to the concept in image domain.
A larger $K$ causes a smaller receptive field, leading to more diverse results.
In particular,
a large $K$ allows only capturing fine-level movements and leads to some unnatural transitions, while a small $K$ leads to overfitting of the original sequence. 
Table~\ref{tab:hyperparameters} shows the global and local patch distance increase as $K$ increases. 
This is because the generated result deviates further from the input sequence when the receptive field is smaller.

\paragraph{Effects of Patch Size $\patchsize$}
The patch size $\patchsize$ defines the temporal length of patches used in the generative matching and blending. 
Patch size controls the receptive field jointly with $K$, and a smaller patch size leads to a smaller receptive field, which creates less coherent result as shown by the increase of global and local patch distance in \Cref{tab:hyperparameters}.

\paragraph{Effects of $\downsamplef$}
The factor $\downsamplef$ controls the step size of transition between stages. A large step size, controlled by a large $\downsamplef$, may result in unstable generation due to big gaps between consecutive scales. On the other hand, a small step size causes unnecessary running time.

\begin{figure}[]
  \centering
  \includegraphics[width=0.95\linewidth]{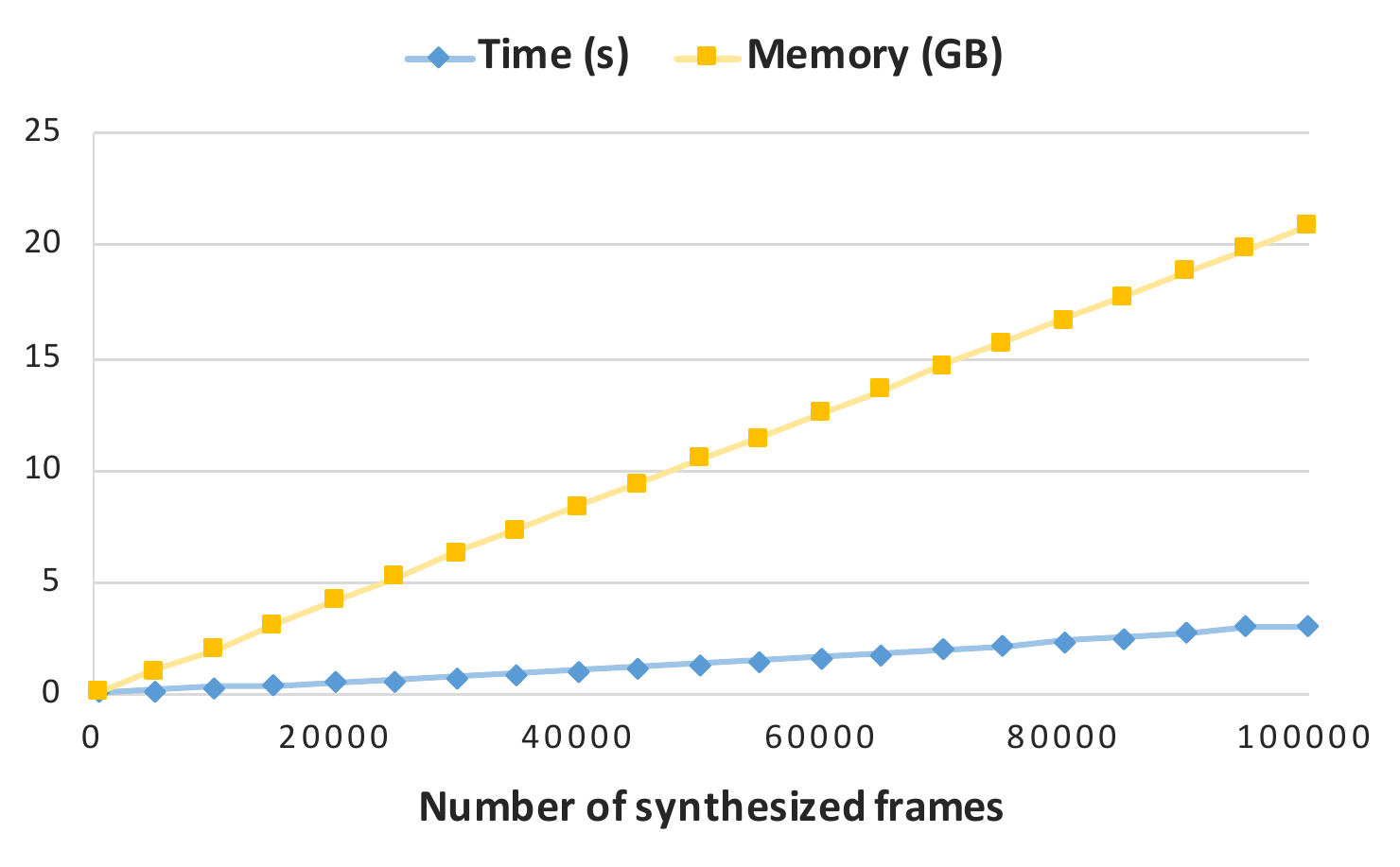}
  \caption{
  Time and memory consumption with respect to increasing numbers of generated frames.
  }
  \label{fig:time_memory}
\end{figure}

\subsection{Time and Memory Consumption}

The memory footprint of the distance metrics described in Section~\ref{sec:patch_generation} increases as the number of generated frames, $\nbframeout$, grows. 
To further investigate the time and memory consumption, we stress-test our method under extreme conditions, where $\exemplar_N$ comprises 522 motion frames of a 65-joint character and we set $\nbframeout$ ranging from 1,000 to 100,000.
These tests are conducted using an NVIDIA V100 GPU equipped with 32GB memory.
Figure~\ref{fig:time_memory} illustrates that both time and memory consumption exhibit a linear growth pattern with respect to the number of generated frames. Owing to the highly paralleled computation of the distance metrics in the GPU,
our method takes only around 3 seconds to synthesize a high-quality sample even consisting of 100,000 frames.

\begin{figure*}[t!]
  \centering
  \includegraphics[width=0.95\linewidth]{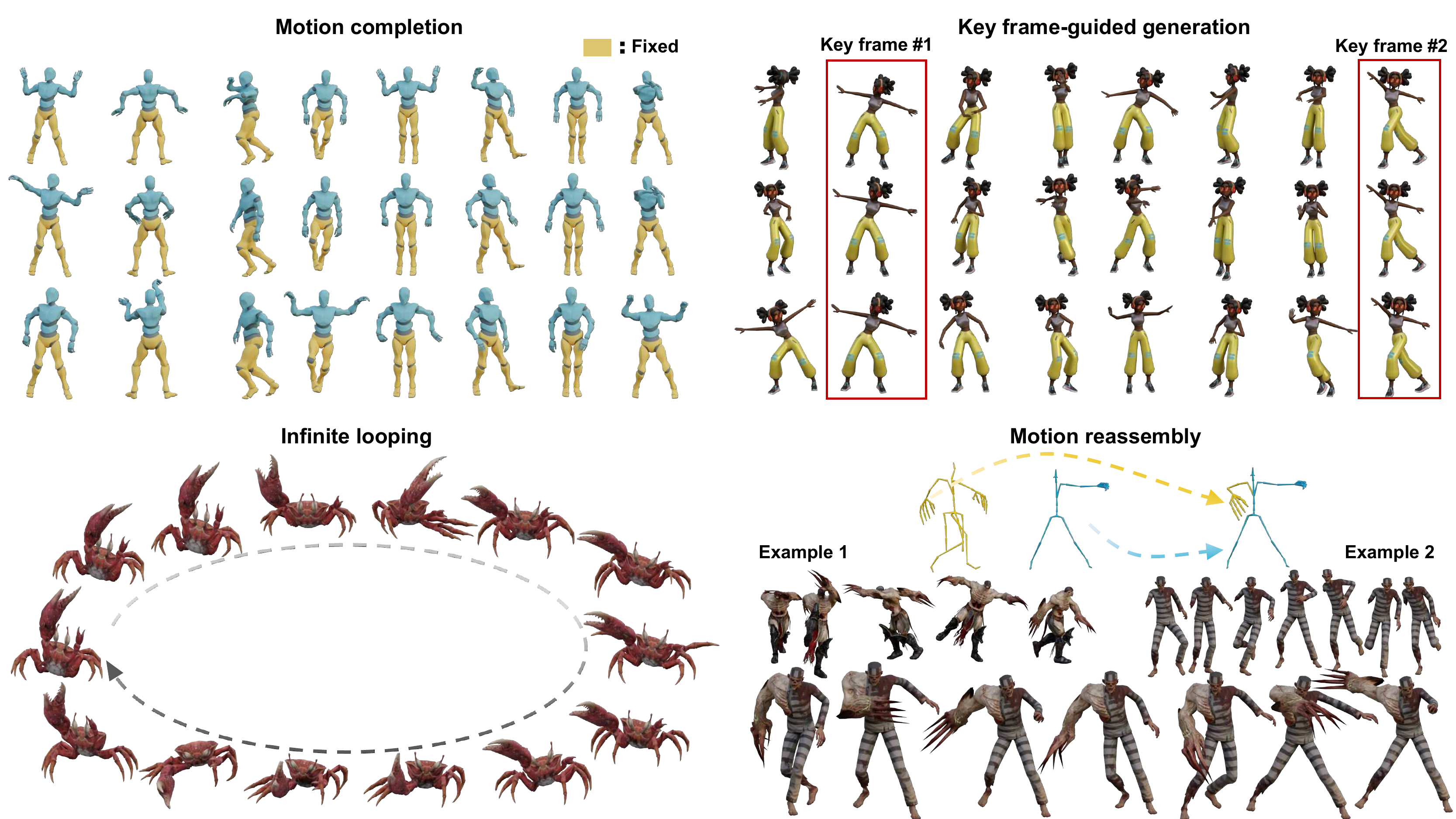}
  \caption{
  Applications.
  (1) Motion completion. Users can provide the lower-body motion (marked in yellow), and our method completes with diverse motions.
  (2) Key frame-guided generation. 
  Given a set of key frames (marked in red boxes), 
  we can generate diverse novel motion sequence that follow the key frame poses.
  (3) Infinite looping. 
  By simply specifying the starting and ending pose to be identical,
  our method can generate a infinitely looping animation, which can be useful in crowd simulation.
  (4) Motion reassembly. 
  Given two motion sequences with heterogeneous skeletons, our method can combine them to form a new creature with coherent and natural motion.
  }
  
  \label{fig:applications}
\end{figure*}

\section{Applications}

In this section, we demonstrate the versatility of our framework
by adapting it to various applications,
such as motion completion,
key frame-guided generation, infinite looping, and motion reassembly.
The results are presented in \Cref{fig:applications}. A more detailed demonstration is available in the accompanying video.

\paragraph{Motion Completion.}
Our framework, which utilizes skeleton-aware motion patch extraction, enables the completion of partial motions that contain only the movement of specific body parts.
For example,
when a lower-body motion sequence $\partialmotion^{\text{lower}}$ is provided,
the upper-body motion can be completed using the example motion.
Specifically,
we build a pyramid for the partial-body motion 
$\{ \partialmotion^{\text{lower}}_1, ...,  \partialmotion^{\text{lower}}_\nbstage \}$,
and the corresponding partial motion in the output $\motionfeatout_s$ is fixed to $\partialmotion^{\text{lower}}_\idstage$ at each stage $\idstage$.
Our framework then automatically synthesizes the movements of the rest by parts, 
completing the partial constraints with a coherent and natural motion.

\paragraph{Key Frame-guided Generation.}
Our method also allows users to manually specify a sparse set of key frames to guide the content of the synthesized motion. 
Our method can then effectively handle these sparse pose constraints distributed throughout the sequence and generate smooth, highly-detailed motion.
Given a set of key frames at the coarsest stage, 
we simply realize it by replacing corresponding frames in $\motionfeatout_1$ with the specified frames,
and fixing them through the whole generation process.
Note that these manually specified key frames should not deviate significantly from the distribution of the poses in the example.
In practice, they can be obtained by simply selecting existing poses in the example, possibly with slight manual modifications by the user.

\paragraph{Infinite Looping.}
Our framework can easily synthesize endless looping motion by fixing the ending pose to be identical to the beginning pose at every stage in the synthesis. This allows for the seamless looping of the entire motion sequence. It can be useful in animation production, such as creating repetitive crowds like spectators cheering outside an arena.

\paragraph{Motion Reassembly.}
As aforementioned in \Cref{sec:more_setting}, our method has the ability to synthesize a Frankenstein. 
We demonstrate an example that stitches the right arm of a monster to a zombie; See~\Cref{fig:applications} and the accompanying video. Note the example sequence of these two characters is different and the zombie does not have any movement in its partially missing right arm, yet our method is still able to successfully synthesize a natural and meaningful motion.

\begin{figure}[t!]
  \centering
  \includegraphics[width=0.96\linewidth]{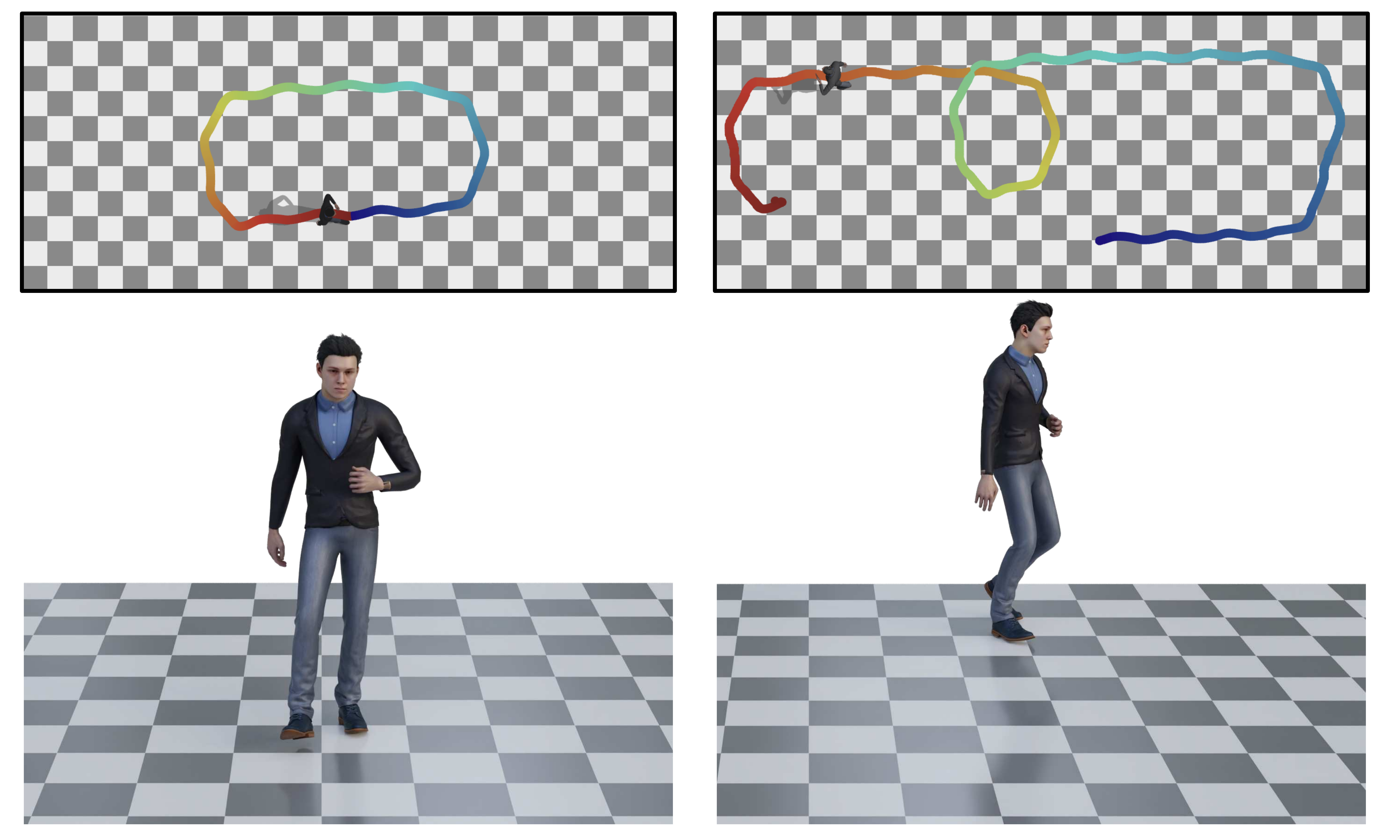}
  \caption{
  Random locomotion generation. 
  Given an example locomotion sequence of a character walking in a circular path (left), we show a high-quality novel motion sequence generated by our method, in which the character walks along a different trajectory (right).}
  \vspace{-15px}
  \label{fig:random_walk}
\end{figure}
\paragraph{Random Locomotion Generation.}
Our method can also generate high-quality novel motion sequences when given a locomotion clip. 
As can be seen in Figure~\ref{fig:random_walk}, 
while the example sequence contains a person walking in a circular path, 
our method can generate novel outputs with a different trajectory (See the difference between the corresponding trajectories at the top row).
More animation results can be found in the accompanying video.

\section{User Interface}
Our framework is general, lightweight, and easy to integrate into many production tools.
For novice users, we build a user-friendly website where users can upload their motion files and then synthesize diverse novel motions with a single click (See the top in~\Cref{fig:blender_addon_and_webdemo}).
We also develop a Blender add-on for professional artists, which seamlessly integrates into their existing workflow as demonstrated at the bottom of \Cref{fig:blender_addon_and_webdemo}.
Note both interfaces can run efficiently on a consumer-level laptop. Please refer to the accompanying video for the results.

\begin{figure}[t!]
  \centering
  \includegraphics[width=0.85\linewidth]{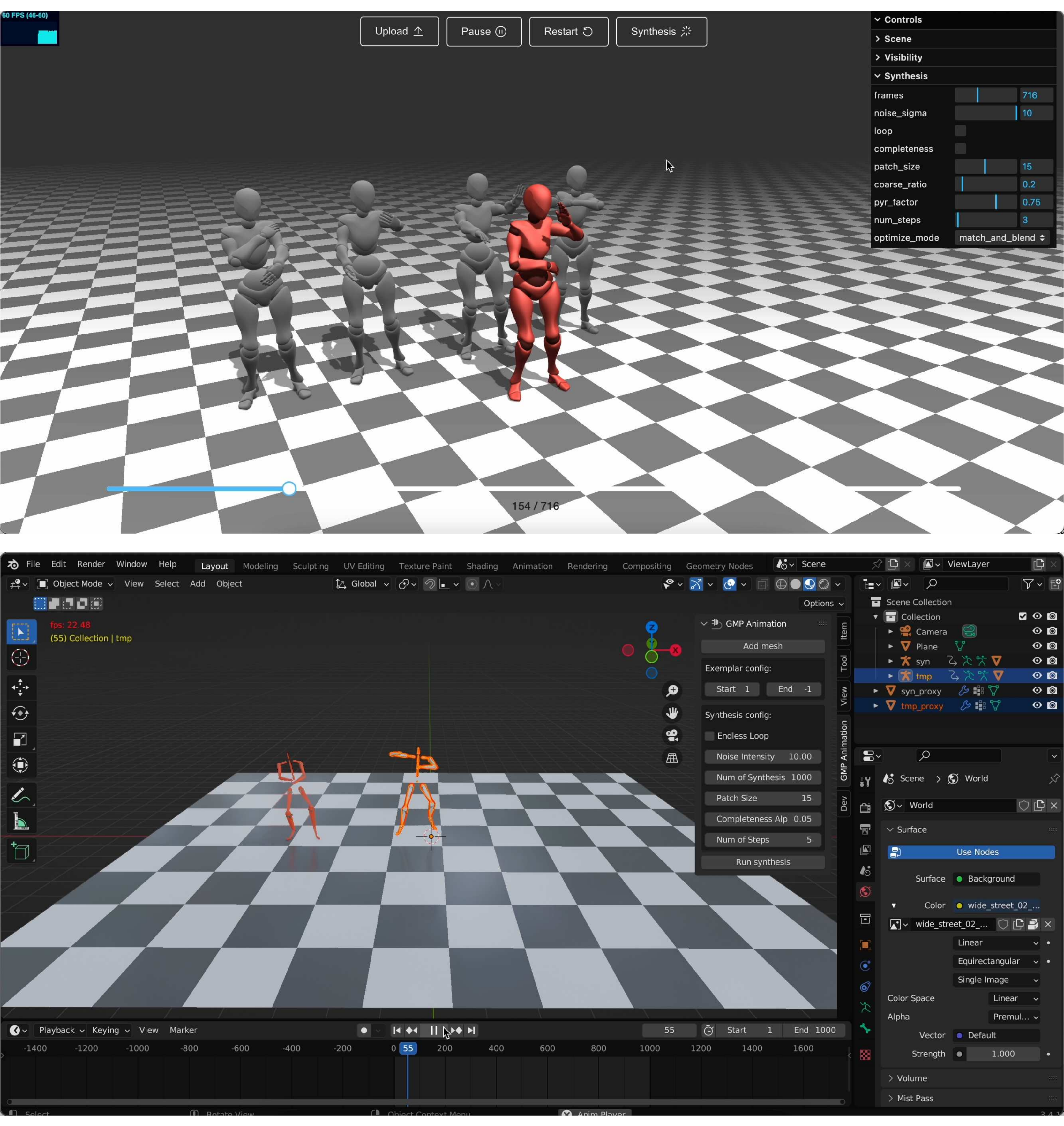}
  \caption{
  Top: screenshot of our web-based interface, characters in grey denote synthesized motions.
  Bottom: screenshot of our Blender add-on, the synthesized motion is highlighted in the middle.
  }
  \vspace{-15px}
  \label{fig:blender_addon_and_webdemo}
\end{figure}
\section{Discussion and Conclusion}
We presented a generative framework for synthesizing diverse motion sequences from only a small set of examples. 
We achieve this via injecting generative capabilities into the industry state-of-the-art technique for character animation \--- motion matching.
As a result,
our framework inherits the training-free nature and superior quality, 
and is able to produce a high-quality sample within just a fraction of a second, even with highly complex and large skeletons.
We demonstrate the
utility of our framework on a variety of applications. 
Despite its advantages, our method in its current form has a few shortcomings:
It uses a discrete patch distribution, whereas GANimator~\cite{ganimator} learns a continuous distribution. 
Therefore,
GANimator can generate novel poses with high likelihood from the learned distribution.
Although the skeleton-aware component can be a remedy,
this capability is missing in our method. 
Nevertheless,
we argue that such generalization can be disadvantageous in motion synthesis, as sequences formed by novel poses often contain visual artifacts such as jittering and incoherence, which are highly noticeable to human eyes.
We prioritized the motion quality at the outset, which led us to  the motion matching approach.

Our method seeks to synthesize as many variants as can be mined from the examples,
rather than struggle to balance quality with novelty of motion.
As a consequence,
although diverse results of our method are shown,
the generative diversity of our method is lower than that of GANimator.
Hence,
a future work direction is to inject the high quality of motion matching into generative neural models, possibly with discrete neural representation learning techniques~\cite{van2017neural}, and thus obtain the best of both worlds.

Our method favors example motions with sufficient intrinsic periodicity, which has been increasingly recognized as an important property of common human motion~\cite{pfnn, starke2022deepphase}, to generate highly diverse novel variations.
It seeks to exploit such patterns in a single example for mining as many coherent variations as possible. 
In extreme cases where the example only involves a single pose change, it may be meaningless to create temporal variations based solely on such input. 
Nonetheless, our skeleton-aware component may introduce variations along the skeletal axis, as evidenced by the asynchronized waving hands in the supplementary video.

Regarding the manual constraints required in the key frame-guided application,
the manually specified key frames cannot differ significantly from those example poses as aforementioned, 
otherwise the generated sequence may not faithfully follow those constraining poses due to the lack of ability to generate completely novel poses as discussed above.

Last, 
our method cannot deal with overly long example sequences, as the normalized similarity matrices grow excessively large.
Adopting approximate nearest neighbors search, such as~\cite{patchmatch}, may help alleviate this issue.

\begin{acks}
We thank the anonymous reviewers for their constructive comments.
This work was supported in part by National Key R\&D Program of China 2022ZD0160801, and the European Research Council (ERC) under the European Union’s Horizon 2020 Research and Innovation Programme (ERC Consolidator Grant, agreement No. 101003104, MYCLOTH).
We would also like to thank Han Liu from Tencent AI Lab for providing the motion data of the avatar Ailing (\Cref{fig:teaser}).
\end{acks}

\bibliographystyle{ACM-Reference-Format}
\bibliography{bibliography.bib}

\end{document}